\documentclass[structabstract]{aa}
\usepackage{graphicx}
\usepackage{txfonts}
\usepackage{paralist}
\begin{document}

   \title{The chaotic solar cycle}

   \subtitle{II. Analysis of cosmogenic $^{10}\rm Be$ data}

   \author{A. Hanslmeier
           \inst{1},
           R. Braj\v sa\inst{2},
           J. \v Calogovi\'c\inst{2},
           B. Vr\v snak\inst{2}
           \and
           D. Ru\v{z}djak\inst{2}
           \and
           F. Steinhilber\inst{3}
           \and
            C.L. MacLeod\inst{4}
           \and 
           \v{Z}. Ivezi\'{c}\inst{4}$^{\rm ,}$\inst{5}
           \and
           I. Skoki\'c\inst{6}
          }

   \institute{Inst. f\"ur Physik, Geophysik Astrophysik und Meteorologie, 
              Univ.-Platz 5, 8010 Graz, Austria\\
              \email{arnold.hanslmeier@uni-graz.at}
              \and
              Hvar Observatory, Faculty of Geodesy, University of Zagreb,
              Ka\v ci\'ceva 26, 10000 Zagreb, Croatia \\
              \email{romanb@geof.hr; jcalogovic@geof.hr; bvrsnak@geof.hr; rdomagoj@geof.hr}
              \and
              Eawag, Swiss Federal Institute of Aquatic Science and Technology,
              \"Uberlandstrasse 133, 8600 D\"ubendorf, Switzerland  \\
              \email{friedhelm.steinhilber@eawag.ch}
              \and
              Department of Astronomy, University of Washington, Box 351580,
              Seattle, WA 98195, USA\\
              \email{cmacleod@astro.washington.edu; ivezic@astro.washington.edu}
              \and
              Department of Physics, University of Zagreb, 
              Bijeni\v cka c. 32, P.P. 331, 10000 Zagreb, Croatia
              \and
              Cybrotech Ltd., Bohinjska 11, 10000 Zagreb, Croatia \\  
              \email{ivica.skokic@cybrotech.hr} \\ }

   \date{Received May 31, 2010; accepted ......}


  \abstract
   {The variations of solar activity over long time intervals using a solar 
    activity reconstruction based on the cosmogenic radionuclide $^{10}$Be 
    measured in polar ice cores are studied.}  
   {The periodicity of the solar activity cycle is studied. The solar activity 
    cycle is governed by a complex dynamo mechanism. Methods of nonlinear dynamics 
    enable us to learn more about the regular and chaotic behavior of solar activity.
    In this work we compare our earlier findings based on $^{14}$C data with the 
    results obtained using $^{10}$Be data.}
   {By applying methods of nonlinear dynamics, the solar activity cycle is studied 
    using solar activity proxies that have been reaching into the past for over 
    9300 years. The complexity of the system is expressed by several parameters
    of nonlinear dynamics, such as embedding dimension or false nearest neighbors, 
    and the method of delay coordinates is applied to the time series. We also fit 
    a damped random walk model, which accurately describes the variability of quasars, 
    to the solar $^{10}$Be data and investigate the corresponding power spectral 
    distribution. The periods in the data series were searched by the Fourier and 
    wavelet analyses.} 
   {The solar activity on the long-term scale is found to be on the edge of chaotic 
    behavior. This can explain the observed intermittent period of longer lasting solar 
    activity minima. Filtering the data by eliminating variations below a certain period 
    (the periods of 380 yr and 57 yr were used) yields a far more regular behavior
    of solar activity. A comparison between the results for the $^{10}$Be data with the 
    $^{14}$C data shows many similarities. Both cosmogenic isotopes are strongly correlated 
    mutually and with solar activity. Finally, we find that a series of damped random walk 
    models provides a good fit to the $^{10}$Be data with a fixed characteristic time scale 
    of 1000 years, which is roughly consistent with the quasi-periods found by the Fourier 
    and wavelet analyses.}
   {The time series of solar activity proxies used here clearly shows that solar activity 
    behaves differently from random data. The unfiltered data exhibit a complex dynamics 
    that becomes more regular when filtering the data. The results indicate  that solar 
    activity proxies are also influenced by other than solar variations and reflect solar 
    activity only on longer time scales.}

   \maketitle

%

\section{Introduction}

The study of the periodic, chaotic, or stochastic nature of solar activity 
requires the analysis of long-lasting time series of solar activity indices.  
Since direct solar activity observations have been available only since the beginning 
of telescopic observations and cover a time span of several hundred years, 
proxies of solar activity have to be used. Such proxies could be observations 
of aurorae, cosmogenic isotopes, growth of certain plants like corals, etc. 
They have to be used to obtain a longer time series of the solar activity.
An overview of these proxies can be found, e.g., in Hanslmeier (2007).

The theory of deterministic chaos has been successfully applied to many areas 
of physics, including geophysics, astrophysics, and meteorology (Peitgen et al. 
1994, 2004; Cvitanovi\'c et al. 2009). Chaotic phenomena in astrophysics and 
cosmology, mainly for dynamics in the solar system and galactic dynamics as well 
as for such applications to cosmology as properties of cosmic microwave background 
radiation, were reviewed by Gurzadyan (2002), while the phenomena showing evidence 
of nonlinear dynamics in solar physics are summarized by Wilson (1994) and by 
Hanslmeier (1997).

The main motivation for this work is to address the question whether 
solar behavior is chaotic or random. The answer to this question 
is important for at least two reasons. First, it has implications for 
the dynamo models, which according to present ideas should describe the
physical processes that govern the observed manifestations of solar 
magnetic activity. Second, this distinction is important for procedures 
of predicting and reconstructing solar activity on different temporal 
scales. 

Chaos is a special kind of complex behavior of dynamic systems described by 
nonlinear equations. The dynamic variables that describe the properties of the 
system and its time evolution are in the nonlinear form, i.e., they have higher 
order terms. Indications of chaotic behavior in solar dynamo models are present in a 
number of cases. The solar activity cycle is modulated by several quasi-periodic 
cycles showing period-doubling characteristics and aperiodic grand minima 
with a characteristic time scale exceeding several tens of cycle periods (Ruzmaikin 
et al. 1992; R\"udiger \& Hollerbach 2004). The impossibility of representing 
long-term sunspot cycle as a periodic process, period-doubling oscillations, and 
positive Kolmogorov entropy found in $^{14}$C data (Gizzatullina et al. 1990) point 
to deterministic chaos (Ruzmaikin et al. 1992). Possible mechanisms include 
\begin{inparaenum}[(i)] 
\item nonlinear back-reaction of the magnetic field visible in a modulation of the differential rotation;
\item stochastic fluctuation of the $\alpha$-effect;
\item variation in the meridional circulation;
\item on-off intermittency due to a threshold field strength for dynamo action.
\end{inparaenum}
Nonlinearity is present in equations of motion, where the Lorenz force is quadratic in 
a magnetic field, and possibly in dynamo equations, where $\alpha$ may also be quadratic 
in a magnetic field (Hoyng 1992; Rosner \& Weiss 1992; Stix 2002; Mestel 2003; 
Ossendrijver 2003; Sch\"ussler \& Schmitt 2004).

This paper is a continuation of our study about the chaotic solar cycle using 
$^{14}$C measurements (Hanslmeier \& Braj\v sa 2010), which is referred to as 
Paper I in this work. In Paper I, references to studies of nonlinear effects 
in the solar activity in theoretical models are given. In addition, fluctuations 
in solar dynamo parameters were considered to explain variability of the solar cycle
(Hoyng 1993; Ossendrijver \& Hoyng 1996; Moss et al. 2008). Chaos and intermittency 
in the solar cycle were reviewed by Spiegel (2009).

Nonlinear dynamics methods have also been applied to predicting solar activity cycles
(see, e.g., Sello 2001; Aguirre et al. 2008; Kitiashvili \& Kosovichev 2008), and one 
has to  distinguish between stochastic and chaotic behavior. The reader is again 
referred to the literature cited in Paper I.

Steinhilber et al. (2008) used the results of Vonmoos et al. (2006) and
reconstructed solar activity for about the past 9300 years using the 
cosmogenic radionuclide $^{10}$Be measured in polar ice cores. In that 
paper, solar activity is expressed as solar modulation potential $\Phi$. 
This $\Phi$ record has been used recently to obtain other records of solar 
activity, such as interplanetary magnetic field (Steinhilber et al. 2010) 
and total solar irradiance (Steinhilber et al. 2009). The abundance depends 
on the intensity of the cosmic ray flux, which can enter the Earth's atmosphere.
The intensity of cosmic rays is lower when the Sun is very active and vice versa.
Thus, there is an anticorrelation between the production of cosmogenic isotopes 
and solar activity. Besides solar activity, the $^{10}\rm Be$ signal is also 
influenced by  geomagnetic field intensity variations and system (climate) 
effects. However, when the data is averaged over 22 or more years, the system
effect component contributes less than 10\% to the total signal (McCracken 2004).
We note that in Vonmoos et al. (2006) and in Steinhilber et al. (2008) the effect 
by geomagnetic field intensity variations has already been considered. However, 
there is large uncertainty in the reconstructions of geomagnetic field intensity. 
In addition, because the geomagnetic field intensity has to be considered when using cosmogenic 
radionuclides (e.g., $^{10}\rm Be$, $^{14}\rm C$), there is also some uncertainty in 
the $\Phi$ reconstruction. To estimate geomagnetic influence, we follow the same 
approach as in Paper I and add random noise with different amplitudes to the $\Phi$ 
record.

While the usual, well-known random walk model has been widely used 
in solar physics (e.g., Leighton 1964; Sheeley et al. 1987; Wang 
\& Sheeley 1994; Hagenaar et al. 1999; Hathaway 2005; Braj\v sa et al. 
2008), this is not the case with the damped random walk (DRW) model. 
In the present work, we use a DRW model to analyze the time 
series of reconstructed solar activity.

In a non-solar context, Kelly et al. (2009) introduced a model
where the optical variability of a given quasar is described by a
DRW. The difference with respect to the
well-known random walk is that an additional self-correcting term
pushes any deviations back towards the mean flux on a time scale
$\tau$. It has been established by Kelly et al.~(2009), Koz{\l}owski et
al.~(2010), and MacLeod et al.~(2010) that a DRW can 
statistically explain the observed light curves of quasars
at an impressive fidelity level (0.01-0.02 mag). The model has
only three free parameters: the mean value of the light curve
($\mu$), the driving amplitude of the stochastic process, and the
damping (or characteristic) time scale $\tau$. The predictions are
only statistical, and the random nature reflects our uncertainty
about the details of the physical processes.

The paper is structured as follows. In Section 2, the data and the applied 
method to analyze the data are described. Section 3 gives the results, and 
in Section 4, we summarize our main results and draw the conclusions.

\section{Data and data analysis}

\subsection{Data}

 \begin{figure}
   \resizebox{\hsize}{!}{\includegraphics{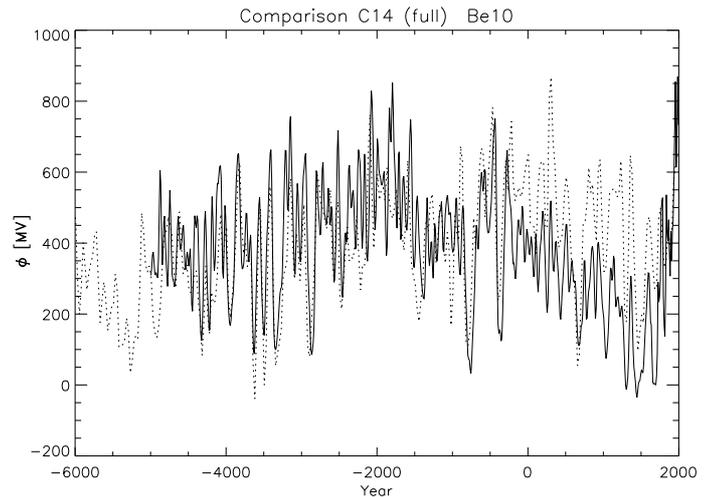}}
   \caption{
      Time series of reconstructed  solar activity ($\Phi$) based on $^{14}\rm C$ (full line) 
      and $^{10}\rm Be$ measurements (dotted line). Years are given in calendar AD years. The 
      data points correspond to centers of the 25-year intervals. Negative values on the $y$-axis 
      are artifacts and are consistent with zero within the error limits due to uncertainty in 
      measuring $^{10}\rm Be$ and in geomagnetic field intensity.
      }
   \label{c14_be}
 \end{figure}

 \begin{figure}
   \resizebox{\hsize}{!}{\includegraphics{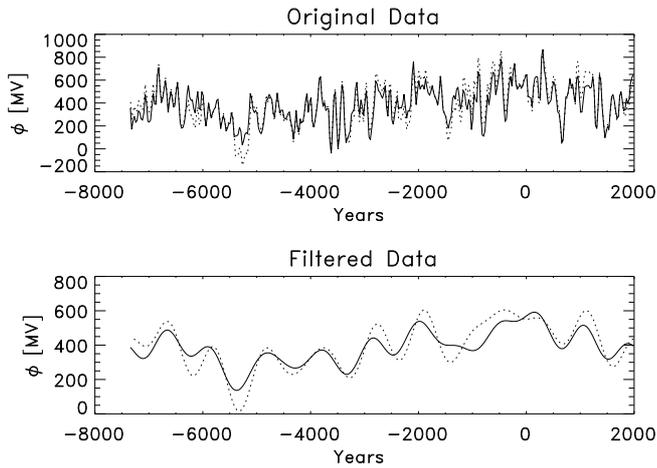}}
   \caption{
      Time series of solar activity ($\Phi$) based on $^{10}\rm Be$ measured in polar ice cores 
      for the past 9300 years. On the upper plot are the original data, while on the lower plot, the 
      Fourier-filtered data are shown. Fluctuations at periods for $t\le 380 \,\rm yr$ (full line) 
      were suppressed by a Fourier filter. To simulate geomagnetic field-strength variation, a 
      random signal of similar amplitude to the original signal and a period of about 760 years 
      were added (same approach as in Paper I) (dotted line).
      }
   \label{15snnr1}
 \end{figure}

  The solar modulation function $\Phi$ during most of the
Holocene (last 9300 years) was reconstructed based on the
cosmogenic $^{10}\rm Be$ data, as described by Vonmoos et al. 
(2006) and by Steinhilber et al. (2008). The data used, which 
were analyzed by different methods of nonlinear dynamics, are 
shown in Figs. \ref{c14_be} and \ref{15snnr1}. The $^{10}\rm Be$ data are limited to 
the last 9300 years due to some missing measurements corresponding 
to the time prior to that period. Also, a climatic change occurred 
at the end of the last glacial period about 11\,700 years ago. 
Such climatic changes could be connected to the change in the 
precipitation rate and thus to the $^{10}\rm Be$ concentration in ice 
(system effects). However, during the Holocene period, there are no 
indications for strong changes in the precipitation source for 
central Greenland (Johnsen et al. 1989; Mayewski et al. 1997).

As stated in the Introduction, the variations seen in the data
should be due to changes in solar activity. However, even though the data are filtered,
there could be still some influence/error due to system effects 
(climate) and uncertainty in the reconstructions of geomagnetic 
field intensity. To test the robustness of our results, we followed 
our approach in Paper I, i.e., stimulating the field by adding 
random data of different amplitudes to the original values and then  
studying how the results changed.

\subsection{Methods of nonlinear dynamics and time series analysis}

We present the calculation of the same parameters that were studied 
in Paper I:
\begin{itemize}
\item mutual information
\item embedding dimension
\item false nearest neighbors
\end{itemize}
For a more detailed description of these parameters, see Paper I or 
the papers by Takens (1980), Sauer et al. (1991), Kennel et al. (1992),  
Kantz \& Schreiber (1997), Rhodes \& Morari (1997), and Schreiber (1999).
The method of delayed coordinates (also described in Paper I) was
used for the $^{10}\rm Be$ data, and a comparison with the $^{14}\rm C$ data is given.

\subsection{Damped random walk}
The time variability is modeled as a stochastic process described by
the exponential covariance matrix
\begin{equation}
        S_{ij} = \sigma^2 \exp(-|t_i-t_j|/\tau)
 \label{eqn:cfunc}
\end{equation}
between times $t_i$ and $t_j$.  As explained by Kelly et al. (2009) and
Koz{\l}owski et al.~(2010), this corresponds to a DRW  with a damping, or
characteristic time scale $\tau$ and a long-term standard
deviation of variability $\sigma$.   
The DRW model used here is more specifically an Ornstein-Uhlenbeck process.
Following Koz{\l}owski et al.~(2010), we model the time series of solar 
activity and estimate the parameters and their uncertainties using the 
method of Press et al.~(1992), its generalization in Rybicki \& Press
(1992), and the fast computational implementation described in
Rybicki \& Press (1995). As in MacLeod et al.~(2010), we express
the long-term variability in terms of the structure function (SF),
where the SF is the root mean square (rms) magnitude difference as a function of the
time lag ($\Delta t$) between measurements. The characteristic
time scale for the SF to reach an asymptotic value
$\rm{SF}_{\infty}$ is the damping time scale, $\tau$. The SF for a
DRW is
\begin{equation}
  SF(\Delta t) = \rm{SF}_{\infty}(1-e^{-|\Delta t|/\tau})^{1/2},
\label{eq:sfdt}
\end{equation}
and the asymptotic value at large $\Delta t$ is
\begin{eqnarray}
  SF(\Delta t >> \tau) \equiv \rm{SF}_{\infty} = \sqrt{2} \sigma.
\label{eq:sfinf}
\end{eqnarray}

In addition to the $\chi^2$ per degree of freedom
($\chi^2/N_{dof}$), information on the goodness of fit is provided
by the parameters $\Delta L_{\rm{noise}}$ and $\Delta L_{\infty}$.
We define $\Delta L_{\rm{noise}} \equiv
\ln{(L_{\rm{best}}/L_{\rm{noise}})}$, which was used in
Koz{\l}owski et al.~(2010) and MacLeod et al.~(2010) to select
quasar light curves that are better described by a DRW than by
pure white noise. Here, $L_{\rm{best}}$ is the likelihood of the
best-fit stochastic model and $L_{\rm{noise}}$ is that for a white
noise solution ($\tau \equiv 0$). We define $\Delta L_{\infty}
\equiv \ln{(L_{\rm{best}}/L_{\infty})}$,
  where $L_{\infty}$ is the likelihood that $\tau \rightarrow \infty$,
  indicating that the length of the time series under consideration 
is too short to accurately
  measure $\tau$.

\subsection{Period analysis}

To find out if there was some periodicity in the analyzed time series,
we applied the Fourier analysis according to the method of Deeming (1975)
and its' later upgraded version of Lenz \& Breger (2005). Furthermore, we 
used the wavelet analysis of Torrence \& Compo (1998). This method made it possible
to identify especially pronounced quasi-periods (see, e.g., Temmer et al. 2004).

\section{Results}

First we give a comparison between the $^{14}\rm C $ and $^{10}\rm Be$
measurements (Fig. \ref{c14_be}). The carbon measurements give the reconstructed
sunspot index; these values were scaled to the $\Phi$ measurements
from the $^{10}\rm Be$ data  as earlier described. It is seen
that the general trend of the data is quite similar, in agreement with
Beer et al. (2007) and Usoskin et al. (2009). The $^{14}\rm C$
data are sampled every ten years, the $^{10}\rm Be$ data every 25
years.

\subsection{Nonlinear dynamics}
In the upper panel of Fig. \ref{15snnr1}, the reconstructed $\Phi$
record is shown. To suppress noise, a Fourier filter of 380 years
was applied to eliminate the shorter time scale fluctuations seen
in the data shown in the lower panel in Fig. \ref{15snnr1}.

 \begin{figure*}
   \sidecaption
   \includegraphics[width=12cm]{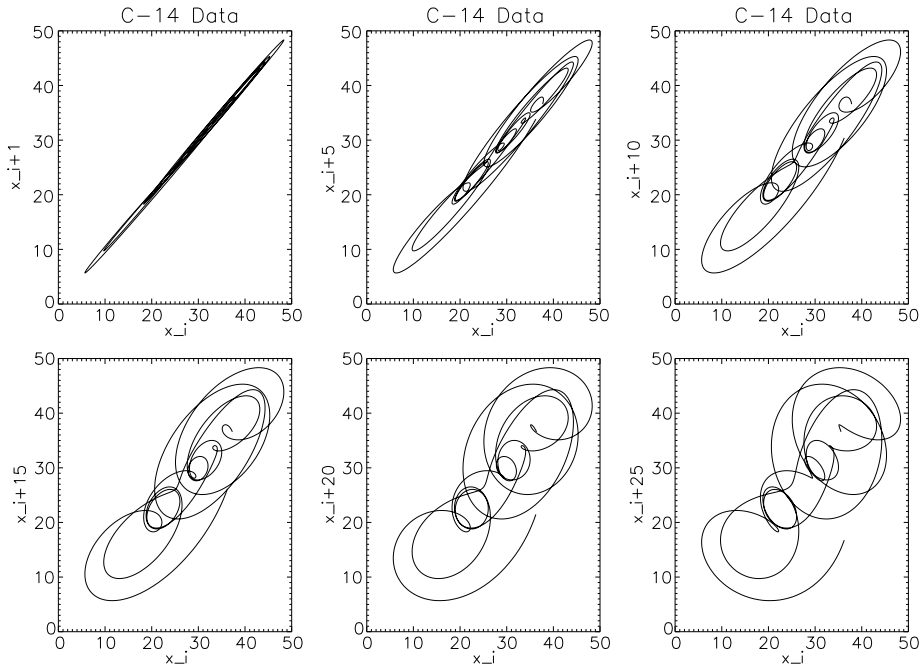}
   \caption{Delayed coordinates for the $^{14}\rm C$
      data. For a discussion, see Sect. 3.1.}
   \label{c14f}
 \end{figure*}

 \begin{figure*}
   \sidecaption
   \includegraphics[width=12cm]{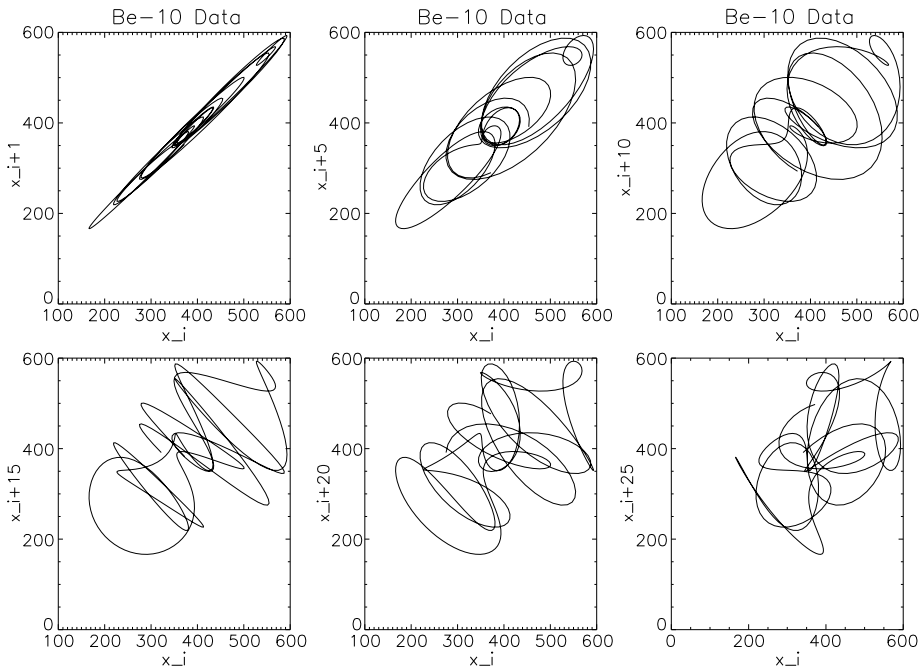}
   \caption{Delayed coordinates for the $^{10}\rm Be$
      data. For a discussion, see Sect. 3.1.}
   \label{Be10f}
 \end{figure*}

We simulated a possible variation in the geomagnetic field to demonstrate 
the robustness of the results. In Fig. \ref{15snnr1} results are also shown for a simulation 
of random geomagnetic variations of a period twice the filtering period of 380 years 
(dotted line). The amplitudes were chosen to simulate a worst case, which means they 
are of comparable amplitude to the original data. These assumptions correspond to 
statements by Solanki et al. (2004, Fig. 3c), who reported a magnetic dipole variation 
of a factor of two at the maximum and of a period that is at least longer than the value 
used for our filtered data.

Next, the delay method was applied. This method is also based on 
the theorem of Takens (1980) and Sauer et al. (1991).
More details on this method can be found in Paper I.
In Figs. \ref{c14f} and \ref{Be10f}, a study of delayed data for
the $^{14}\rm C$ and the $^{10}\rm Be$ is given. The delay values
were taken as $\Delta t=1, 5, 10, 15, 25$, which means the data
values $x_i$ are plotted versus $x_{i\Delta t}$. We note that the
value of $\Delta t=1$ means a time gap of 10 yr in the case of the
$^{14}\rm C$ data and 25 yr in the case of the $^{10}\rm Be$ data.
Since we only want to demonstrate the chaotic or nonchaotic
behavior for the given data sets, this difference does not have any
significance here. The results shown are for the filtered data, which
distinctly show some regular behavior. The results given in Figs. 
\ref{c14f} and \ref{Be10f} clearly demonstrated that the dynamics 
becomes more complex when going to variations on a shorter timescale.
The behavior of the attractor given by the delay coordinates does not 
show a significant difference with respect to the original filtered data.
By such a filter, variations below a time scale of 57 years were eliminated 
(not shown here).

\begin{figure}
   \resizebox{\hsize}{!}{\includegraphics{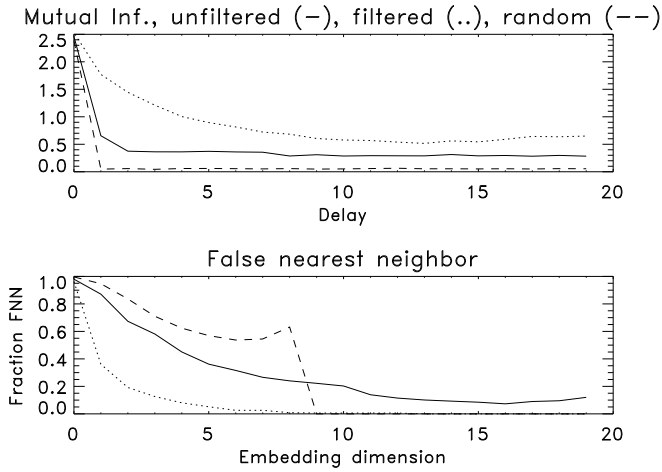}}
   \caption{
      Mutual information  as a function of delay  and false nearest neighbors  
      for unfiltered data (full line), filtered data (dotted line), and random data 
      (dashed line).
      }
   \label{mut}
 \end{figure}

In Fig. \ref{mut} the mutual information is shown as a function of delay 
(upper panel) and the false nearest neighbors (lower panel) for the the 
different datasets: unfiltered data (full line), filtered data (dotted 
line), and random data (dashed line). For the filtered data, the mutual 
information curve decreases less steeply than for the original data.  
For the random data, the graph immediately declines to zero.

\begin{figure}
   \resizebox{\hsize}{!}{\includegraphics{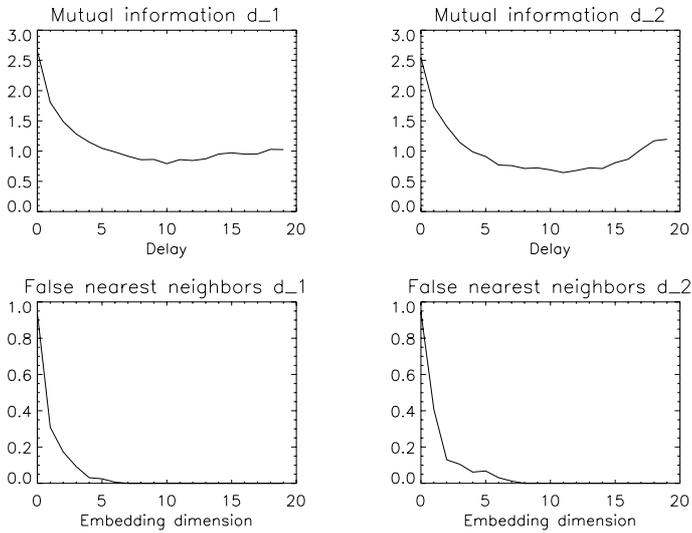}}
   \caption{
      Mutual information as a function of delay and false nearest neighbors  
      for two halves of the filtered $^{10}\rm Be$ dataset. Left-handed (right-handed) 
      plots refer to the first (second) subset of the data.
      }
   \label{plot_mut_fnn}
 \end{figure}

\begin{figure}
   \resizebox{\hsize}{!}{\includegraphics{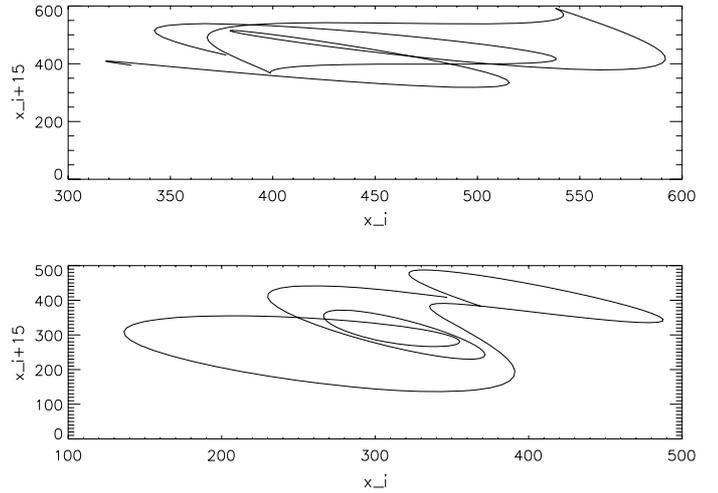}}
   \caption{
      Delayed coordinates for two halves of the $^{10}\rm Be$ dataset. Upper 
      (lower) plot referes to the first (second) subset of the data.
      }
   \label{plot_del5}
 \end{figure}

To investigate what effect the length of the time series could have on the 
results, we split the $^{10}\rm Be$ data into two halves and compare the 
results for each dataset (Figs. \ref{plot_mut_fnn} and \ref{plot_del5}). 
The results concerning false nearest neighbor and mutual information were 
found to be quite similar. Therefore, our results are not influenced by the 
limits of the available time series.

\subsection{Damped random walk}
\begin{figure}
   \resizebox{\hsize}{!}{\includegraphics{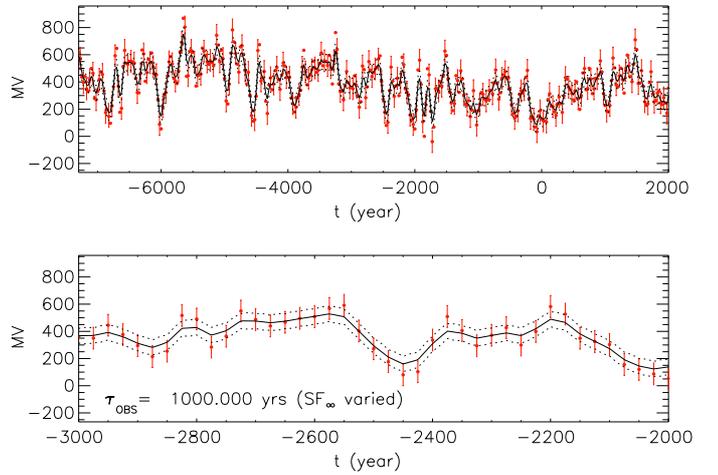} }
   \caption{The top panel shows the entire time series for the solar $^{10}\rm Be$ data. 
      Data points with error bars show the observed data. The solid line shows 
      the weighted average of all consistent DRW models with $\tau=1000$ years 
      (SF$_{\infty}$ is allowed to vary from model to model). 
      Dotted lines show the $\pm 1\sigma$ range of these possible stochastic 
      models about the average. The bottom panel shows a zoomed-in segment of the time series.
      }
   \label{fig:DRW}
 \end{figure}

Table~\ref{tab:DRWresults} lists the best-fit DRW parameters for
the solar $^{10}\rm Be$ data, assuming an uncertainty of 80 megavolts (MV) for 
all data points. A relative likelihood of $\Delta L_{noise} = 29$ 
indicates that the best-fit DRW model provides a better fit to the 
$^{10}$Be data than pure white noise. However, a single DRW model 
cannot accurately reproduce the data and yields a reduced $\chi^2$ 
of 1.9. A series of DRW models with $\tau$ fixed to 1000 years and 
varying SF$_{\infty}$ produces an average model that is in perfect 
agreement with the data to within the adopted errors, whereas 
a series of DRW models with varying $\tau$ and fixed SF$_{\infty} = 165$~MV 
cannot reproduce the data. Therefore, the best-fit characteristic 
time scale of 1000 years seems to be well constrained. However, 
we note that the time scale of 1000 years has a large uncertainty when 
measured using a DRW analysis due to the limited light curve length 
for the solar data. The best-fit DRW model gives $1\sigma$ Bayesian 
upper and lower limits on $\tau$ of $10^{4.4}$ and $10^{2.9}$ years, 
respectively. The data and best-fit DRW models are shown in Fig.~\ref{fig:DRW}. 

\begin{table}
  \centering 
  \caption{Best-fit DRW parameters for solar $^{10}\rm Be$ data.}
  \begin{tabular}{c c}
  \hline\hline
    $\mu$ (MV) & 390.68 \\
    $\tau$ (years) & 1000 \\
    $\rm{SF}_{\infty}$ (MV) & 165.52 \\
    $\chi^2/N_{dof}$ & 1.9 \\
    $\Delta L_{\rm{noise}}$ & 29 \\
    $\Delta L_{\infty}$ & 192 \\
  \hline
  \end{tabular}
  \label{tab:DRWresults}
\end{table}

 \begin{figure}
   \resizebox{\hsize}{!}{\includegraphics{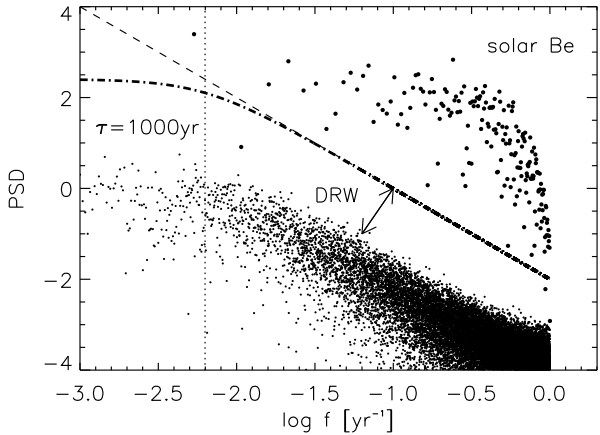}}
   \caption{
      PSD comparison of the solar $^{10}$Be data (large dots) 
      and a DRW model (smaller dots) with $\tau = 1000$ years (indicated by the dotted 
      line). A DRW process has a spectral index of $-2$ on short time scales (shown by 
      the dashed line), which is shallower than that for the $^{10}$Be data on short 
      time scales. The dotted-dashed line shows the average PSD for many DRW models with 
      $\tau=1000$ years.
   }
   \label{psd_drw}
 \end{figure}

The inability of a single DRW process to describe the solar $^{10}$Be 
data is due to an inaccurate power spectral distribution (PSD) description 
on short time scales. In Fig. \ref{psd_drw}, the PSD for the solar $^{10}$Be 
data is compared to the PSD for a DRW model. The $x$-axis is log frequency 
in [1/years].  The $y$ axis has arbitrary units.  The smaller dots show the 
PSD for a DRW process with a characteristic time scale arbitrarily set to 
$\tau = 10^{2.7}$ years (indicated by the vertical dotted line), and the 
larger dots show the PSD for the solar $^{10}\rm Be$ data. The dashed line 
shows a spectral index of -2, which corresponds to a DRW process on time scales 
shorter than $\tau$.  However, the solar data show a steeper PSD slope than a 
DRW at frequencies larger than about $10^{-0.1}$ yr$^{-1}$. The dotted-dashed 
line shows the average PSD for many DRW models with $\tau=1000$ years, which is 
still too shallow to accurately describe the solar data.  This discrepancy between 
the data and DRW models can be seen in Fig. \ref{fig:DRW}, where some of the observed points 
fall outside of the $1\sigma$ range of model light curves (shown as dotted lines), 
but the data are still consistent with the models, given their uncertainty of 80 MV. 
In other words, the data generally show a larger scatter on short time scales with 
respect to the DRW model.

\subsection{Period analysis}

\begin{figure}
   \resizebox{\hsize}{!}{\includegraphics{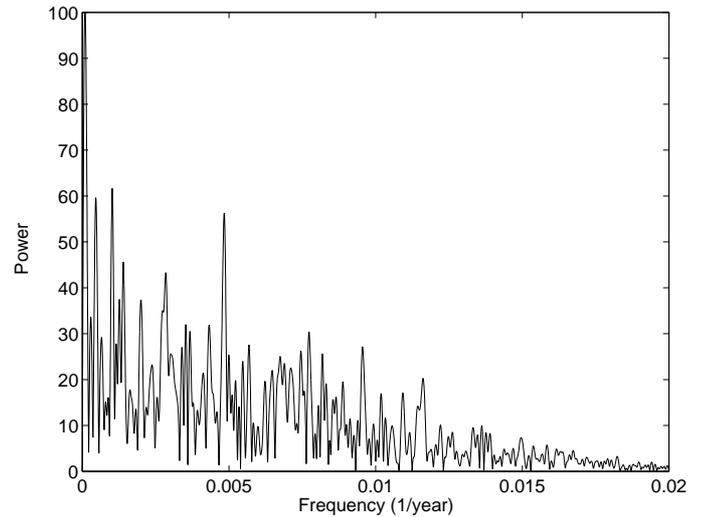}}
   \caption{Fourier spectrum for solar $^{10}\rm Be$ data.}
   \label{Be04power}
 \end{figure}

\begin{figure*}
   \centering
   \includegraphics[width=17cm]{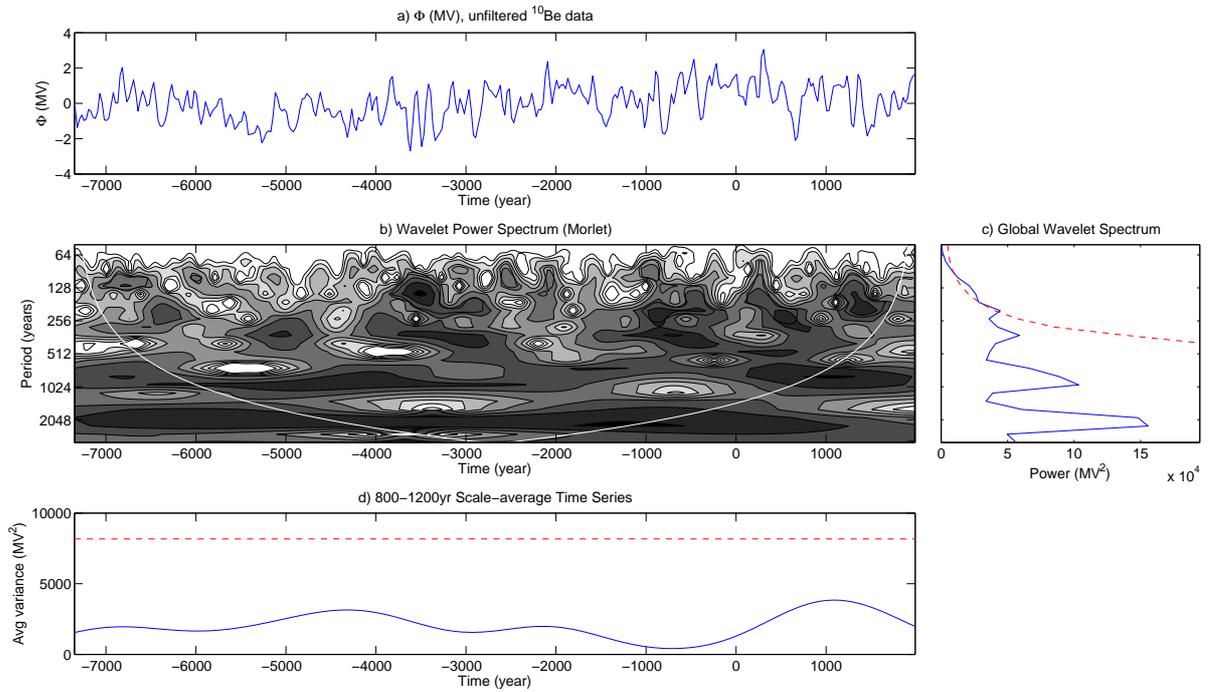}
   \caption{Wavelet analysis of the solar $^{10}\rm Be$ data using the Morlet wavelet. 
     (a) Time series of solar modulation function ($\Phi$) anomalies from unfiltered $^{10}\rm Be$ data. 
     (b) Wavelet power spectrum: darker contours indicate the higher values of wavelet power. 
         White thin line is the cone of influence, below which edge effects become important.  
     (c) Global wavelet power spectrum, i.e., the temporal mean wavelet power at each period. 
         Dashed line is the 95\% confidence level. 
     (d) Scale-averaged time series for the period band 800-–1200 years. Dashed line is the 95\% 
         confidence level. For more information about wavelet analysis, see Torrence \& Compo (1998).
     }
   \label{wavelet_10Be3}
 \end{figure*}

The solar $^{10}\rm Be$ time series was also analyzed by Period04 software
(Lenz \& Breger 2005). Table~\ref{tab:Be04periods} lists several strongest
quasi-periods as seen in Fourier spectrum (Fig. \ref{Be04power}) with a dominant 
approximatively 1000-year cycle. The wavelet analysis results obtained are 
presented in Fig \ref{wavelet_10Be3}. We see that by applying the wavelet analysis 
a quasi-period of about 1000 yr can also be identified, but the significance is 
not very high and varies during the analyzed time interval.

\begin{table}
  \centering 
  \caption{Strongest periods from Fig. \ref{Be04power} for solar $^{10}\rm Be$ 
    data as calculated by Period04 software.}
  \begin{tabular}{r r}
  \hline\hline
  No. & Period (years) \\
  \hline
    1. & 10970 \\
    2. &  976 \\
    3. & 2169 \\
    4. &  207 \\
    5. &  715 \\
    6. &  351 \\
    7. &  790 \\
    8. &  500 \\
    9. &  365 \\
    10. & 3454 \\
  \hline
  \end{tabular}
  \label{tab:Be04periods}
\end{table}

\section{Discussion and conclusions}

We applied standard methods of nonlinear dynamics (the time series 
analysis including calculation of the embedding dimension and delay, 
the false nearest neighbor estimation, and the investigation of the 
mutual information; for details see Paper I) to study the behavior 
of solar activity. It must be stressed that the data consist of 
averaged values, where the averaging was done over a 25-year interval. 
Therefore, the results are only indicative of longer term variations 
and do not include and represent the 11-year solar activity cycle because 
of the short-term noise and limited sampling rate in the $^{10}\rm Be$ 
data. Hence, the conclusions are valid for longer time scale modulations 
of solar activity. To make a further distinction between shorter time 
scale and longer time scale variations, we applied a Fourier low-pass filtering.

The mutual information and false nearest neighbor methods applied give 
insight into the complexity of the underlying system. From this information,
we estimated the embedding dimension $m$. In the case of the unfiltered data, 
we obtained the value $m=15$. At this value, the number of false nearest 
neighbors becomes very small (Fig. \ref{mut}).  In the case of filtered data, this 
value is lower, while for the elimination of fluctuations with $t\le 380 \,\rm yr$, 
the value is about five. Both values are similar to the values obtained in Paper 
I. The complexity of the system therefore strongly increases on shorter 
time scales.

Also, we repeated the method of delay coordinates for different delay values.
The results for the $^{10}$Be data are similar to the ones obtained with the
$^{14}$C data (see Paper I). The topological structure is quite complex when 
considering the unfiltered original data. The structure, however, becomes fairly
simple when considering the filtered data, eliminating $t\le 380 \,\rm yr$ 
fluctuations.This can be interpreted that solar activity proxies seem to exhibit 
a more regular and predictable behavior when variations at larger time scales 
are considered. These larger time scales are well above 100 years.

Concerning the methods of nonlinear dynamics, we have analyzed the time series, 
estimated the false nearest neighbors, and calculated the mutual information. 
The mutual information is a generalization of the autocorrelation function. 
Also, the method of delays and the calculation of the embedding dimension $m$ 
were applied. The embedding dimension can help to answer the question whether 
the topological structure of a system in a phase space is preserved by transformation 
(mapping). The topological structure is preserved if neighbors are mapped into neighbors,
and it is not preserved if this is not the case. Moreover, it is well known that the Lyapunov 
exponents determine the evolution of a dynamical system.

The topological structure of a system is preserved if $m > m_0$, and it is not preserved 
if $m < m_0$, where $m_0$ is the minimal embedding dimension for a time series. 
In that last case, the neighbors are mapped into false neighbors.

For the data used in the present analysis, we can estimate the value of 
$m_0$ from the upper panel of Fig. \ref{mut} (mutual information as a function of delay). 
All curves become fairly constant at the y-value of about one or less. As a result, $m_0 = 1$. 
Now from the lower panel of Fig. \ref{mut} (fraction of false nearest neighbors as a function of $m$), we can estimate 
$m$ for the unfiltered and filtered data. For the unfiltered data, $m = 15$, and for 
filtered data, $m$ has a smaller value of 7. So we can conclude that the criterion $m > m_0$ holds for both 
the filtered and the unfiltered data, implying 
that the topological structure is preserved, i.e., that the neighbors are fairly well 
mapped into neighbors after the transformation.

The application of false nearest neighbors and embedding dimension in the astrophysical (and especially
in the solar context) are discussed in detail by Regev (2006) and by Gilmore and 
Letellier (2007). 
In order to test our analysis against the dependence on the number of data points used,
we split the $^{10}\rm Be$ data into two halves and compared the results for
each subset. As the results are quite similar, we conclude that they are not
influenced by those limits.

Finally, we fit a DRW model, which accurately describes the
variability of quasars, to the solar $^{10}$Be data. In this way, we find that 
a series of DRW models provides a good fit to the $^{10}$Be
data with a characteristic time scale of 1000 years.

The best-fit $\tau$ from the DRW analysis (1000 years) seems to be real, although
the power spectrum of the solar $^{10}$Be data has a different shape than that for a DRW
process (the latter has a shallower power-law slope).  Although the DRW process is
linear and stochastic, this does not necessarily mean that the solar $^{10}$Be data are not
consistent with a chaotic, nonlinear, and deterministic process, even if the data
were perfectly described by the DRW model.   Stochastic processes are commonly used
to approximate an astrophysical time series that is truly chaotic in nature, and
our analysis in Section 2.3. may indeed represent a similar approximation.

In addition, we note that Ma (2007) found a 1000-year cycle in solar 
activity by investigating long-term fluctuations of reconstructed 
sunspot number series. It is interesting that a quasi-period of
about 1000 years was also identified in the present work using Fourier 
and wavelet analyses, though there were equally strong quasi-periods at 207 and 2169 years.
Concerning the influence of geomagnetic field variations, one can estimate 
that variations on a time scale of the order of 1 ky are of solar origin, 
but  longer periodicities $>3\,\rm ky$ may be caused by magnetic modulation.

Our results on the stochastic and chaotic properties 
of solar activity based on $^{10}$Be data are quite similar to the results 
obtained with the $^{14}$C time series discussed in Paper I, confirming the 
findings of Beer et al. (2007) and of Usoskin et al. (2009). This demonstrates 
the robustness of our method and indicates that such an analysis tool is suited 
to study the complex time behavior of such systems. Both cosmogenic isotopes 
are strongly correlated mutually and with solar activity.

The cosmogenic radionuclides $^{14}$C and $^{10}$Be as long-term indices for 
solar activity were studied by Beer (2000a, 2000b) and Steinhilber et al. (2008). 
They stress that these data are produced in a similar way, but that their geochemical 
behavior is different. The $^{10}$Be production rate (0.018 cm$^{-2}{\rm s}^{-1}$) 
is more than 100 times lower than that of $^{14}$C because it is only produced by 
high-energy spallation processes, where $^{14}$C is produced by thermal neutrons 
interacting with nitrogen. After production, $^{10}$Be gets attached to aerosols 
and is immediately removed from the atmosphere after about one year. It is then
transported to the ground, where it is archived, e.g., in polar ice. In contrast, $^{14}$C 
after production forms $^{14}$CO$_2$ and is included in the global carbon cycle, i.e., 
it is exchanged between the reservoirs of CO$_2$, such as atmosphere, biosphere, and the 
oceans. Consequently, it has much longer residence times, depending on the 
reservoir (atmosphere: ten years, biosphere: 60 years, ocean: 1000 years).

Thus, the $^{14}$C signal (called $\Delta$ $^{14}$C) is not the $^{14}$C production rate 
at one timepoint, but reflects the $^{14}$C production integrated over several millennia.
In addition, one has to consider that the halflife of $^{14}$C is about 5730 years, 
which is of the same order as the  exchange processes between atmosphere, biosphere, and 
the ocean. As a result, the carbon cycle and the radioactive decay of $^{14}$C have to be 
considered in order to get the $^{14}$C production rate out from the $\Delta$ $^{14}$C. Both have 
been considered  in Solanki et al. (2004), whose record was the basis for Paper I.

The carbon cycle itself has no influence on the transport of $^{10}$Be to the polar ice, 
and thus we can conclude here that the similarities we found between these two 
radionuclides indicate that the dominant signal in the data is the production signal due 
to solar activity and geomagnetic field intensity variations rather than a climate signal.
Our findings agree with the study by Beer et al. (2007), who applied principal component 
analysis to both radionuclides. They found that most of the variation is described 
by the first principal component, which reflects production.
In a continuation of this work, we plan to perform the Hurst analysis of the $^{14}$C and
$^{10}$Be data.

\begin{acknowledgements}
  The research leading to the results presented in this paper received partial funding 
  from the European Community's Seventh Framework Programme (FP7/2007-2013) under grant agreements 
  nos. 218816 (SOTERIA) and 263252 (COMESEP) and from the Alexander von Humboldt Foundation.
  F. Steinhilber acknowledges financial support by NCCR Climate - Swiss climate research.
  \v Z. Ivezi\'c acknowledges support by the Croatian National Science Foundation grant 
  O-1548-2009. The authors also acknowledge the support from the Austrian-Croatian Bilateral 
  Scientific Project (2010/11) for financing the exchange of scientists and would like to thank 
  B. Kelly for helpful insight regarding the DRW process as well as J. Beer and the anonymous referee for 
  helpful comments and suggestions. 
\end{acknowledgements}

\end{document}